\documentclass{aa}  

\usepackage{graphicx}
\usepackage{txfonts}
\begin{document}

   \title{Kinematic, Elemental and Structural Dependences on Metallicity in the Galactic Bulge}

   \author{Haoyang Liu
          \inst{1}
          \and
          Cuihua Du\inst{1,} \thanks{Corresponding author: 
          	ducuihua@ucas.ac.cn}
          \and
          Zhongcheng Li\inst{1}
          \and
          Jian Zhang\inst{1}
          \and
          Mingji Deng\inst{1}
          }

   \institute{College of Astronomy and Space Sciences, University of Chinese Academy of Sciences, Beijing 100049, P.R. China\\
             }

   \date{}

 
  \abstract
   {The Galactic Bulge is known to have multiple stellar populations according to the peaks of metallicity distribution function (MDF). These populations show varying properties of kinematic, elemental and spatial distributions.}
   {The study aims to investigate metallicity-dependent variations in the kinematic properties, elemental abundances, and spatial distributions of Galactic Bulge stars and distinguishes these populations.}
   {We selected Bulge stars from APOGEE DR17 cross-matched with astrometric data from \textit{Gaia} DR3. Bulge stars were divided into sub-samples with line-of-sight velocity dispersion analyzed and the peaks of MDF were detected by both Gaussian Mixture Models (GMM) and \texttt{scipy.signal.find\_peaks}. Orbital parameters (e.g., actions) were calculated via orbital integration, and the kinematic dependence on metallicity was investigated. GMM is also conducted to kinematically distinguish the metal-poor and metal-rich populations. Analyses were put on the Bulge stars (including retrograde stars), their elemental abundances, and the [Mg/Mn]-[Al/Fe] plane to investigate potential accreted components. Finally, the shapes (X-shaped/boxy) of Bulge stars with different metallicities were analyzed through least-squares fitting based on the analytical Bulge models.}
   {By studying the kinematic, elemental and structural dependences on metallicity for Bulge stars, our findings are concluded as follows: 1. Six peaks are detected in the Bulge MDF, encompassing values reported in previous studies, suggesting a complex composition of Bulge populations. 2. An inversion relationship is well-observed in metal-rich sub-samples, while absent in metal-poor sub-samples. 3. Metal-poor populations exhibit larger dispersions than metal-rich stars (which is also revealed by GMM decomposition), suggesting that metal-rich stars are kinematically coherent. 4. Retrograde stars are confined to $\sim1$ kpc of the Galactic center, with their relative fraction decreasing at higher [Fe/H] -- a trend potentially linked to the ``spin-up'' process of Galactic disks. 5. Metal-rich Bulge stars with [Al/Fe] $<-0.15$ are likely associated with from disk accreted substructure, while all elemental planes exhibit bimodality but Na abundances rise monotonically with metallicity. 6. In general, stars with all metallicities support a boxy profile.}
   {}


   \maketitle
%

\section{Introduction}\label{sec:intro}
The bulges and bars are commonly found in disk galaxies, with at least $\sim$45\% of disk galaxies exhibiting these features \citep{L2000}. There are two main types of bulges: classical bulges and pseudo-bulges. The classical bulges are formed through hierarchical mergers \citep{Wyse1997} with predominant random motions of stars \citep{Fisher2016} and they have a surface brightness profile with Sersic index n$>2$ and kinematics similar to elliptical galaxies \citep{Kormendy2004,A2005}. By contrast, the pseudo-bulges could have formed by disk and bar instabilities with a larger timescale \citep{Combes1981,Wyse1997}, or by a high-redshift starburst (see Okamoto \citeyear{O2013} for an alternative view). 

Previous studies have shown that the Galactic Bulge is a pseudo-bulge in boxy/peanut (B/P) shape with red clump giants (RCG) star counts method \citep{Stanek1994,Ba2005,NessL2016}. An X-shaped component in the Bulge at higher latitude is also identified using RCG star counts \citep{McWilliam2010,Nataf2010,wegg2013} and simulations \citep{Gardner2014,Li2015,Abbott2017}. However, there are arguments that the X-shaped structure could be an artifact, as RCG do not exhibit a distinct narrow peak in their luminosity function, which is attributed to observational evidence for an X-shape in the density \citep{Lopez2016,Lopez2019}. The morphology of the Galactic Bulge is also related to stellar ages and metallicity. \citet{semczuk2022} used the Auriga simulation to find that younger stars exhibited a more pronounced B/P shape and X-shaped component when projected side-on, compared to older stars. \citet{A2017} found that the metal-poor stars form a flattened spheroidal-like component, while the metal-rich stars show a clear X-shape when viewed side-on in their simulation. 

The presence of multiple stellar populations in the Galactic bulge leads to several peaks in the metallicity distribution function (MDF). The MDF has been studied using various tracers such as RGB, RCG, M giants and dwarfs \citep{Zoccali2008,U2012,Hill2011,Ness2013a,Rich2012,Bensby2017}. Notably, several peaks have been discovered in the metal-poor range ($-0.8<$[Fe/H]$<0$) in previous studies, while only two peaks at $\sim0.10$ and $\sim0.30$ in the metal-rich region ([Fe/H] $>0$), possibly suggesting a more complex origin for those metal-poor stars and a relatively coherent origin for the metal-rich stars. Using a limited sample of metal-poor stars, \citet{Lucey2019} suggested these stars were neither halo interlopers nor accreted populations, based on their large dispersions in both elemental abundances and $\alpha$-element enhancement. 
However, the metal-poor Bulge may still keep records about the proto-Galaxy \citep{Rix2022,Horta2024} or possible accreted substructures, known as Kraken/Koala/Heracles \citep{Kraken,Forbes2020,Horta2021} that are within the inner Galaxy. By contrast, the origin of the metal-rich stars in the Bulge has not been thoroughly explored, and the reason why the extremely metal-rich stars are confined within $\sim$ 1.5 kpc of the Galactic center remains unclear \citep{Rix2024}.

The kinematic properties of the Galactic Bulge also exhibit a strong dependence on metallicity. Along the minor axis of the Bulge, metal-poor stars have larger dispersions of radial velocity and line-of-sight velocity than the metal-rich stars at $|b|>$ 4° as revealed by giant stars \citep{Rich1990,M1996}. An inverted trend could also be observed at lower latitudes where the metal-rich stars show bar-like high velocity dispersions \citep{B2014,B2016}.  This situation differs from the Galactic disk, where metal-rich young disk stars are less perturbed by the gravitational potential and exhibit lower velocity dispersions \citep{Garzon24}. \citet{Boin2024} conducted an N-body simulation in which metal-rich stars were placed in a thin disk, while metal-poor stars were assigned to a thick disk. Their simulation demonstrated that metal-rich stars were more easily trapped by bar instability, successfully reproducing the observed trend at low latitudes and suggesting a disk origin for the Bulge. This is not surprising, as early studies have indicated the chemical similarities between the local thick disk and the Galactic Bulge \citep{Mel2008,Alves2010}, indicating that the Bulge could form from the buckling instability of the bar and migration of disk stars \citep{Erwin2016,lian2021,Molero2024}. Stellar migration is driven by changes in angular momentum and epicyclic amplitude, processes known as ``churning" and ``blurring" \citep[e.g., see][]{S2009}. Across latitudes of $|b|<5$°, the metal-poor stars with [Fe/H]$<-1$ show the slowest rotation and flattest velocity dispersion, while the metal-rich counterparts have the fastest rotation and highest dispersions with $0<l<35$° \citep{Ness2016}, which suggests a boxy profile of the Bulge.

The bimodal distribution of Bulge stars in the [Mg/Fe]-[Fe/H] plane has been primarily discovered and studied by \citet{Rojas2019}, who showed that the stars in the metal-rich and metal-poor modes of the Bulge's MDF represent distinct evolutionary sequences that overlap at super-solar metallicity. The downward trend of [Mg/Fe] at super-solar metallicity was first observed by \citet{Johnson2014} and \citet{Gonzalez2015}, where the knee is between $-0.6$ and $-0.4$ of metallicity ([Fe/H])\citep{Friaca2017,Bensby2017}. The bimodality of other $\alpha$ elements such as O, Si and Ca (which also includes [C/N] abundances) is also observed in \citet{Queiroz2021}, suggesting a discontinuous star formation path. For iron-peak element such as Mn, [Mn/Fe] exhibits an upward trend with increasing [Fe/H] \citep{McWilliam2003,Barbuy2013,Sch2017}, which is consistent with the chemical evolution model in the solar neighborhood with Asymptotic Giant Branch (AGB) stars  and super AGB stars included \citep{Kobayashi2020}. Furthermore, the [Mn/O]-[Fe/H] distribution for the Bulge stars overlaps with those of thick and thin disk stars \citep{Barbuy2013}, suggesting that its evolutionary paths may resemble both disks, potentially explaining the $\alpha$-element bimodality or indicating possible disk origin for some Bulge stars. Regarding the odd-z elements Al and Na, Al abundances behave like an $\alpha$ element, exhibiting a bimodal pattern in both observations \citep{McWilliam2016} and chemical evolution models \citep{Kobayashi2006}. In contrast, the decreasing trend of Na at higher metallicity is not well predicted by \citet{Kobayashi2006}.

The kinematics, chemical distributions, and structural characteristics of different populations in the Galactic Bulge may differ significantly as indicated above, suggesting distinct origins and evolutionary paths for stars of varying metallicities. Therefore, we study the kinematic, elemental, and structural dependences on metallicity in the Galactic Bulge to explore the underlying properties that have not been fully diagnosed before. For kinematics of Bulge stars, we include the actions (i.e., $J_r$, $J_z$, and $J_{\phi}$) which could provide deeper dynamical insights as they are integrals-of-motion under a symmetric gravitational potential. This paper is structured as follows: Section~\ref{data} gives a brief introduction on data selection and derivation of orbital parameters. The results are presented in Section~\ref{result}. The discussion and summary are given in Section~\ref{summary}.

\section{data} \label{data}
In this work, we use the elemental abundances from the Apache Point Observatory Galactic Evolution Experiment (APOGEE) data release 17 \citep{apogeedr17} and proper motions and positions from \textit{Gaia} data release 3 \citep{GAIA2016,Gaia2023}. We apply the following cuts for data quality: EXTRATARG = 0; SNR$>70$; log $g<3.6$ dex; $3500<\text{TEFF}<5500$ K; VERR $<$ 10 km/s; FE\_H\_FLAG = 0; [Fe/H]$>-0.8$; \text{dist\_error} / \text{dist} $<$ 0.2. The first five conditions aim to select giant stars with high signal-to-noise ratios, precise heliocentric velocities, and accurate metallicity estimates. The metallicity cut is used to remove the major contamination from Galactic stellar halo interlopers \citep{Feuillet2022}. 
Studies have used apocenter radius $R_{\text{ap}}>3.5$ kpc \citep{Han2025} to remove possible interlopers. The threshold proposed by \citet{Lucey2022} could exclude the interlopers from the metal-weak thick disk (MWTD). The percentages of stars with $R_{\text{ap}}<3.5$ kpc and $R_{\text{ap}}<6$ kpc in the subsequent Bulge sample are 73.2\% and 98.2\% respectively. \citet{Yan2022} revealed that MWTD contributed significantly only when [Fe/H] $<-0.8$ in the inner region. Therefore, our sample contains negligible fraction from MWTD interlopers. Besides, interlopers from the inner stellar halo usually have $R_{\text{ap}}\gtrsim6$ kpc \citep[e.g.,][]{Deason2018}, thus the sample contains negligible interlopers from the MWTD and halo by this metallicity cut. Since only $\sim5\%$ stars with [Fe/H]$<-1$ in the Galactic Bulge, this selection will not affect the general study of Bulge populations \citep{NessF2016}. The last condition is to constrain the precision of distances, which are taken from the astroNN value-added catalog \citep{Leung2019}. The astroNN neural network incorporates $K_s$ extinctions, assuming no uncertainty in the extinction correction due to the high precision of APOGEE stellar extinction estimates. Moreover, extinction could alternatively be inferred jointly from the spectra and multi-band photometry, so the distances for Bulge stars are trustworthy.

\begin{figure*}
    \begin{center}
        \includegraphics[width=0.9\textwidth]{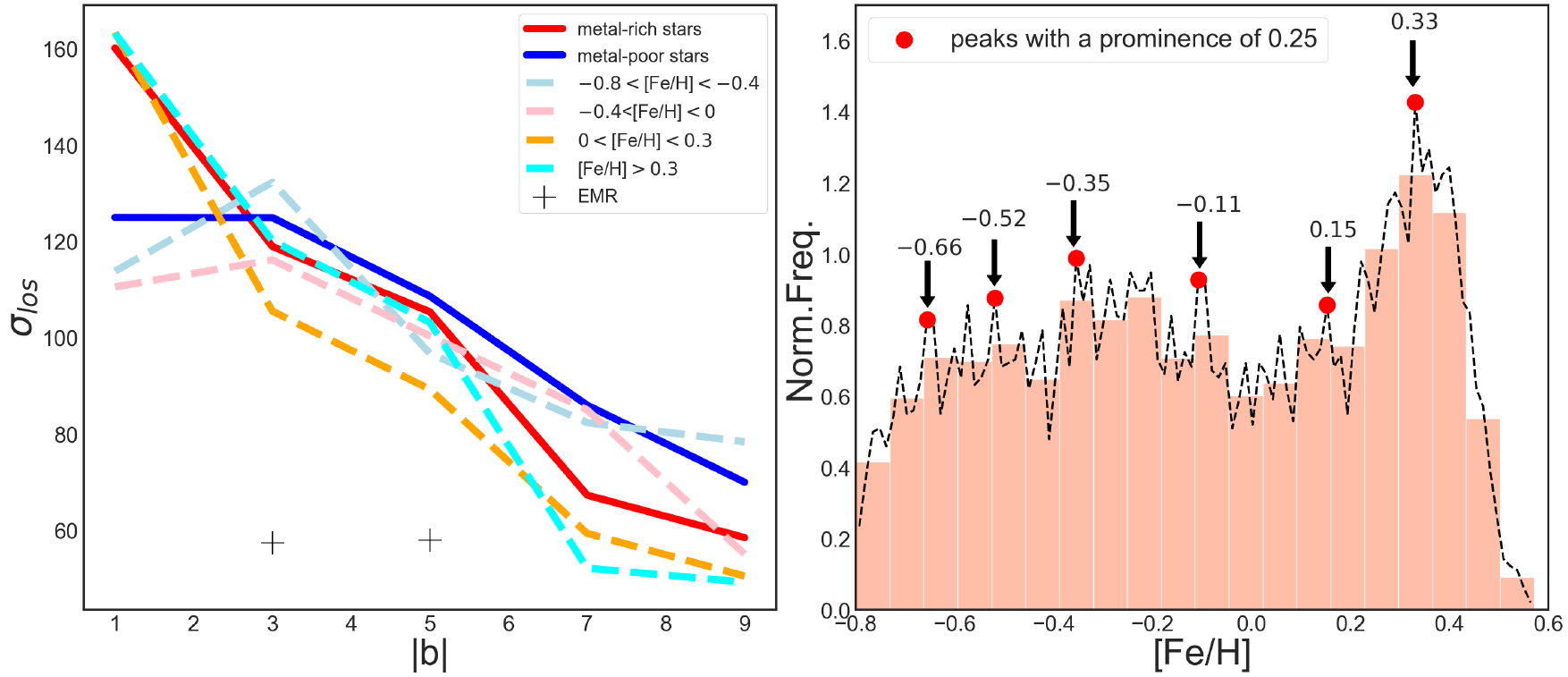}  
        \caption{Left panel: line-of-sight velocity dispersions (in units of km/s) across $0<|b|<10^\circ$ along the minor axis ($|l|<2^\circ$) of the Bulge. The solid red and blue lines represent metal-rich and metal-poor stars, respectively. The dashed lines represent the metal-rich and metal-poor sub-samples, while the black cross indicates EMR stars. Right panel: the MDF of the Bulge. Six peaks have been identified with a prominence of 0.25, as indicated by the black arrows.}
        \label{dispersion}
    \end{center}
\end{figure*}

\begin{table*}[h]
    \caption{Comparisons between the peaks of the MDF from this study and those from previous literature where the parentheses indicate the methods}
    \centering
    \begin{tabular}{|c|c|c|c|}
        \hline
        Literature & Peaks & Number  \\ \hline
        this work    & $-0.66,-0.52,-0.35,-0.11,0.15,0.33$ (\texttt{scipy.signal.find\_peaks})    & 7156       \\ \hline
        this work    & $-0.67,-0.49,-0.28,-0.08,0.16,0.36$ (GMM)  & 7156        \\ \hline
        \citet{U2012}    & $-0.6,0.3$  (corrected MDF using bootstrapping)  & 401       \\ \hline
        \citet{Zoccali2008}    & $-0.7,-0.3,0.3$ (Gaussians; one zone model)   & 650      \\ \hline
        \citet{Johnson2013}    & $-0.44,-0.29,-0.08$ (0.1 dex binning)   & 264      \\ \hline
        \citet{Ness2013a} & $-0.70\sim-0.67$, $-0.27\sim-0.22$,$0.13\sim0.16$ (Bayesian; GMM) & \textbackslash \\ \hline
        \citet{Sch2017}    & $-0.29,0.30$ (GMM)  & 269       \\ \hline
        \citet{Hill2011}    & $-0.30,0.32$ (kernel estimation)   & 219       \\ \hline
        \citet{Rojas2017} & $-0.36\pm0.08$, $0.40\pm0.05$ (GMM) & 1583 \\ \hline
        \citet{Zoccali2018} & $<-0.8$, $-0.4$, 0.3 (GMM) & 5500 \\ \hline
        \citet{Wylie2021} & $-0.50\sim-0.40$, $\sim0.30$ (corrected MDF)& \textbackslash \\ \hline
        \citet{John2022} & $\sim-0.30$, $\sim0.20$ (fields binning) & \textbackslash \\ \hline
    \end{tabular}
    \label{mdf}
\end{table*}

For the selection of Bulge stars, we require  cylindrical radius R $<3.5$ kpc \citep{Ness2013} and $|z|<1$ kpc, resulting in a total sample of 7,156 stars. It should be noted that we do not impose any additional quality cuts when analyzing kinematics and spatial distributions of the Bulge stars to maintain a slightly larger sample size. However, for the analysis of elemental abundances, we require that ASPCAPFLAG = 0 to ensure the abundances quality, which results in a sample of 5,772 stars. 

For the derivation of orbital parameters, we integrate the orbits backward over a period of 5 Gyr using the galpy package \citep{bovy2015}. Here the Local Standard of Rest (LSR) velocity $V_{\text{LSR}}$ is 232 km/s and the cylindrical distance of the Sun is 8.21 kpc \citep{McMillan2017}. The peculiar motion of the Sun is $[U_{\sun}, V_{\sun}, W_{\sun}] = [11.1, 12.24, 7.25]$ km/s \citep{Sch2010} and the Sun is located at 20.8 pc above the Galactic mid-plane \citep{Bennett2019}. We adopt the axisymmetric gravitational potential \texttt{MWPotential2014} \citep{bovy2015} to obtain the conserved quantities: the radial action $J_r$, the vertical action $J_z$, the azimuthal action $J_{\phi}$ (which is equal to $L_z$), and other orbital parameters such as eccentricity ($e$) and the maximum height above the mid-plane ($Z_{\text{max}}$). The potential adopted here includes a spherical bulge and dark matter halo, and a Miyamoto-Nagai disk \citep[see detailed parameters in Table 1 of][]{bovy2015}. We also calculate the circularity ($\eta$) of the orbits, which is defined as $|L_z|/L_c(E)$, where $L_c(E)$ is the angular momentum with given energy ($E$).

\section{Results}\label{result}
\subsection{Line-of-sight Velocity Dispersions and MDF of the Galactic Bulge}
In this work, we define metal-rich stars with [Fe/H]$>0$ and metal-poor stars with $-0.8<$[Fe/H]$<0$ in later analyses. We divide the metal-poor and metal-rich stars into two sub-samples: $-0.8<$[Fe/H]$<-0.4$ and $-0.4<$[Fe/H]$<0$ for metal-poor stars; $0<$[Fe/H]$<0.3$ and [Fe/H]$>0.3$ for metal-rich stars. We also adopt the definition of extremely metal-rich stars (EMR) with [Fe/H]$>0.45$ in \citet{Rix2024}. We calculate the dispersions of line-of-sight velocity along the minor axis ($|l|<2$°) of the Bulge in five $|b|$ bins within the range 0$<|b|<10$°. As shown in the left panel of Figure~\ref{dispersion}, the metal-poor stars have larger line-of-sight velocity dispersions than metal-rich stars when $|b|\gtrsim3$°, which is in good agreement with previous studies \citep{Bab2010,B2014,Ness2013,Rojas2014}.

It is also observed that the metal-rich sub-samples exhibit an inverted trend at higher latitude $|b| \sim 6.5^\circ$. The two metal-rich sub-samples exhibit similar dispersions, suggesting that they may be drawn from a single distribution. The p-values from the Kolmogorov-Smirnov test and F test are 1.000 and 0.551, respectively, failing to reject the null hypothesis. This suggests that the two sub-samples likely originate from the same distribution. However, this result is expected, given that the metal-rich stars exhibit a peak around 0.3 (as seen in the later MDF analysis). Using this 0.3 threshold to divide the sub-samples inevitably leads to significant overlap in their distributions.

However, the inverted trend does not strictly hold for the metal-poor sub-samples, as stars with $-0.8 < \text{[Fe/H]} < -0.4$ still have larger dispersions than those with $-0.4 < \text{[Fe/H]} < 0$ at $|b|\lesssim3$°. A similar pattern observed in metal-poor stars is also presented in \citet{Khoperskov2024}, who used an orbit superposition approach. However, in their analysis, stars with $-1.0 < \text{[Fe/H]} < -0.5$ exhibit larger dispersions than those with $-0.5 < \text{[Fe/H]} < 0$, nearly across all regions with $|b| < 15^\circ$. The dispersion differences suggest that the metal-poor and metal-rich stars have a distinct origin. In the N-body simulation of \citet{Boin2024}, the metal-poor stars ([Fe/H] $< -0.5$) and metal-intermediate stars ($-0.5 < \text{[Fe/H]} < 0$) still exhibit the inverted trend at low latitudes. This indicates that the metal-poor stars origin from the thick disk and the metal-rich stars from thin disk cannot fully account for the observations. Interestingly, EMR stars have very low line-of-sight velocity dispersions $\sim60$ km/s (only 2 bins for $2<|b|<6^\circ$ due to the small sample size), which indicates that EMR stars have more coherent kinematics or nearly zero mean line-of-sight velocity in the inner Bulge (see Figure 5 of Rix et al.\citeyear{Rix2024}).

The peaks of MDF in the Galactic Bulge are identified by adopting \texttt{scipy.signal.find\_peaks} function and Gaussian Mixture Model (GMM), respectively. The \texttt{scipy.signal.find\_peaks} function successfully identified six peaks in the MDF with a prominence of 0.25, which are close to those reported in previous studies using RGB stars and Red Clump (RC) stars. We applied GMM by minimizing the Bayesian Information Criterion (BIC) value to find the optimal number of Gaussian components. Six components were also identified by GMM, although there are slight shifts in the resulting values. It is worth emphasizing that the results of GMM may lack statistical significance, as its prior assumption that each component follows a Gaussian distribution is unfavorable for potentially skewed distributions. However, the MDF of the sample shows no significant tail, making the use of GMM a valid simplification. Moreover, the discovered peaks are similar, indicating that these components are securely fitted. The validity of GMM for unbiased distributions is further supported by previous studies \citet{Ness2013a} and \citet{Zoccali2018}.

From Table~\ref{mdf}, the metal-rich peak at $\sim$ 0.3 is the easiest to identify among previous studies, as it has the highest prominence of $\sim$1.20, while the metal-rich peak at $0.15$ only has a prominence of $\sim$0.31. Metal-poor stars generally exhibit multiple peaks, whereas metal-rich stars typically exhibit only one or two peaks. This suggests that metal-poor stars have more complex origins, while metal-rich stars may share a common origin. As suggested by \citet{Matteucci2019}, metal-rich populations may originate from the inner disk due to the secular evolution of the bar, while metal-poor populations could be linked to the chemically complex inner and outer bulge regions \citep{Lucey2022}.

In Table~\ref{mdf}, most studies indicate a metal-rich peak around 0.3 \citep{Zoccali2008,Zoccali2018,U2012,Sch2017,Wylie2021}, 0.32 for \citet{Hill2011}, as well as the values of 0.33 and 0.36 obtained in this work. It is worth noting that a less metal-rich peak has been detected by both \citet{Ness2013a} (0.13) and this work (0.15 and 0.16). \citet{Bensby2013} and \citet{Bensby2017} used microlensed dwarf and detected a peak at [Fe/H]$\sim0.08$ and $\sim0.12$, respectively. \citet{Ness2013a} attributed the peak to the instability-driven of bar/bulge formation from the thin disk, and it is more prominent at lower latitude ($|b|=-5^{\circ}$). Since $\sim71\%$ of the sample have $|b|\lesssim5^{\circ}$, it is reasonable that this peak is detected. Regarding the peak at $-0.11/-0.08$ in this work, a similar feature was only detected in \citet{Gonzalez2011}, where the median [Fe/H] ratio in the field (+5, -3) was $-0.08$. However, since no other studies have confirmed this result, the peak may be spurious. As for the three metal-poor peaks detected in this work, they exhibit values consistent with those reported in previous studies. For the possible contamination from disk stars here, \citet{Rojas2020} have found a spatial limit of R = 3.5 kpc could well separate the rotation-supported disk and bar+pressure supported Bulge kinematically. Moreover, \citet{Boin2024} selected Bulge stars in a spatial way to analyze the Bugle kinematics, their results were consistent with previous studies and their simulation. Therefore, the disk contamination do not have a crucial affect in the analysis.

\begin{figure*}
    \begin{center}
        \includegraphics[width=0.9\textwidth]{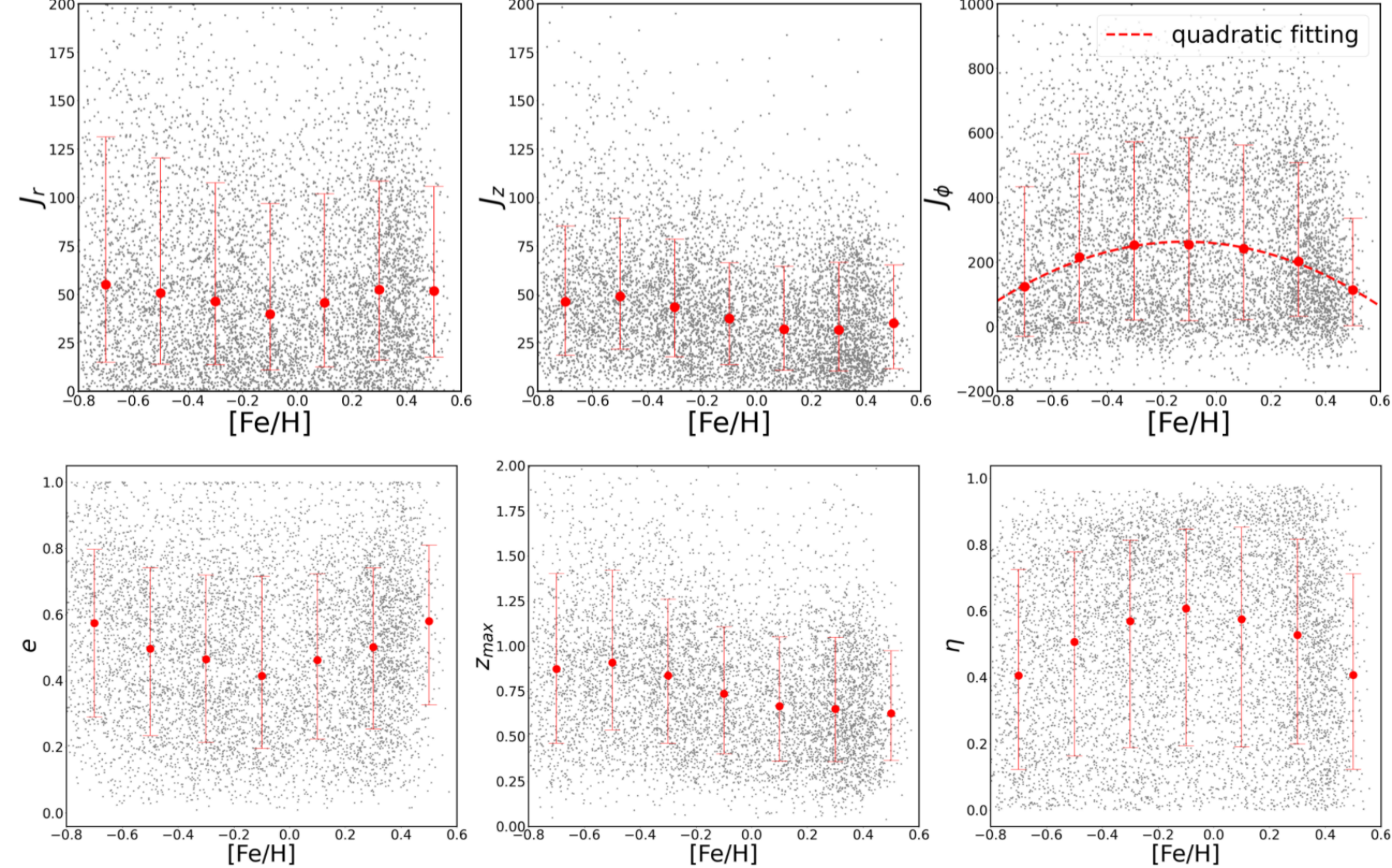}  
        \caption{The kinematic dependence on metallicity in the Galactic Bulge.  The red dot and error bar indicate the median values, as well as the 16th and 84th percentiles for each equal-width [Fe/H] bin. Top row: the distributions in action-metallicity plane. Note that the median values of $J_{\phi}$ in each bin could be well fitted by a quadratic function. Bottom row: the distributions in $e$/$Z_{\text{max}}$/$\eta$ planes.}
        \label{kinematic}
    \end{center}
\end{figure*}

\begin{figure*}
    \begin{center}
        \includegraphics[width=0.8\textwidth]{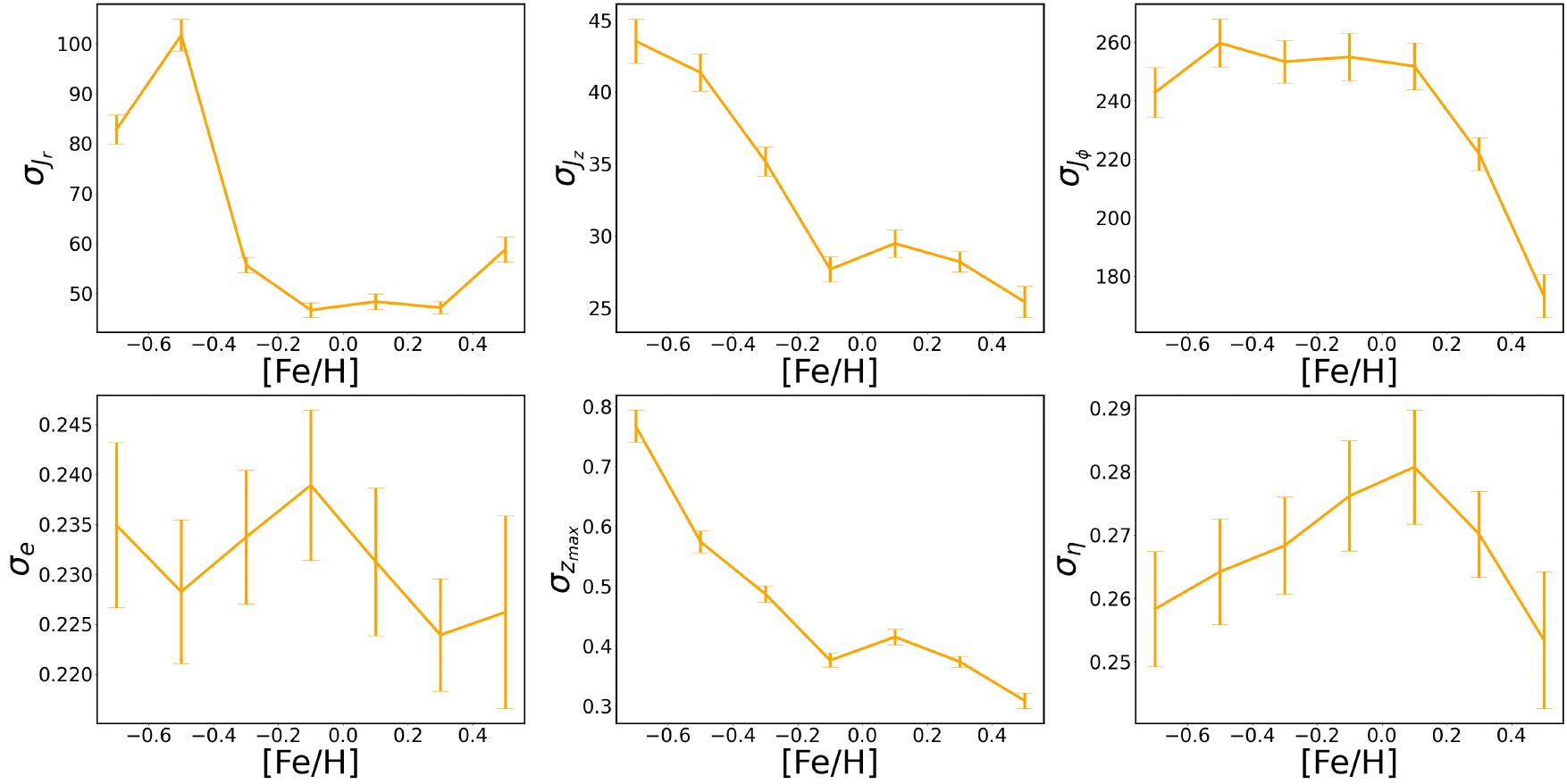}  
        \caption{The dependences of the action dispersions (defined as standard deviations) and orbital dispersions on metallicity is examined. The action dispersions of metal-poor stars are larger than those of metal-rich stars, suggesting that metal-poor stars may have multiple dynamical origins. The error bar indicates the standard error, which is given by $\sigma/\sqrt{N-1}$.}
        \label{sigma}
    \end{center}
\end{figure*}

\begin{figure*}
    \begin{center}
        \includegraphics[width=0.95\textwidth]{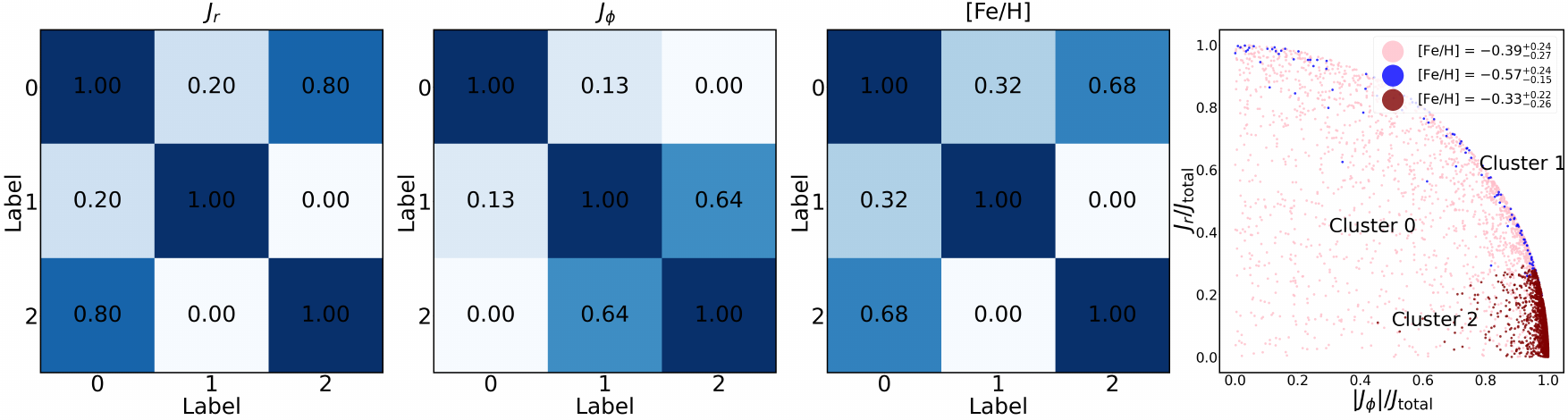}  
        \caption{The heatmap illustrates the similarity of the three clusters based on the distributions of $J_r$, $J_z$, and [Fe/H]. The rightmost panel shows the distributions of the three clusters in $J_r / J_{\text{total}}$-$|J_{\phi}| / J_{\text{total}}$ plane, with the median, 16th percentile, and 84th percentile of [Fe/H] for each cluster indicated in the legend. Cluster 0 has a wide span in the plane and could potentially be clustered into smaller groups when additional quantities are considered. Cluster 1 is the most metal-poor and is supported by both radial and rotational motion. Cluster 2 is the least metal-poor and is mainly supported by rotational motion.}
        \label{heatmap}
    \end{center}
\end{figure*}
\subsection{Kinematic Dependence on Metallicity in the Galactic Bulge}

Figure~\ref{kinematic} illustrates the kinematic dependence on metallicity in terms of actions ($J_r$, $J_z$ and $J_{\phi}$), $e$, $Z_{\text{max}}$ and circularity $\eta$. The trends of $e$, $Z_{\text{max}}$ and $\eta$ with respect to metallicity are similar to those of actions ($J_r$, $J_z$ and $J_{\phi}$), as they are corresponding physical quantities. From the middle column of Figure~\ref{kinematic}, there is a downward trend of $J_z$ with increasing metallicity, indicating that metal-rich stars are vertically colder than metal-poor stars, allowing them to be more easily centered in the mid-plane (also see Ness et al.\citeyear{Ness2016}). Interestingly, \( J_r \) (corresponding to \( e \)) exhibits a downward trend in the metal-poor range \((-0.8 < \text{[Fe/H]} < 0)\) and then an upward trend in the metal-rich range \((\text{[Fe/H]} > 0)\). Additionally, there is a clear inverse trend in the \( J_{\phi} \) - \text{[Fe/H]} plane. The more metal-poor stars are typically older and  experienced more gravitational perturbation, thus leading to higher vertical velocity dispersions (i.e., $J_z$) and more eccentric orbits (i.e., $J_r$ and $e$). This is counterintuitive compared to the trends exhibited by metal-rich stars; however, the more metal-rich stars are generally centered in the inner bulge, and they tend to have, on average (the decreasing $L_z$ with increasing [Fe/H]), more eccentric orbits than metal-poor stars. However, it is challenging to explain how these metal-rich stars could be centered within a small galactocentric distance if they entered the bulge region through radial migration, which necessitates further development of the dynamical model in the future.

To further study the correlation between kinematics and metallicity, we also examine the dispersions of action variables and orbital parameters. Figure~\ref{sigma} shows that metal-poor stars clearly exhibit higher dispersions of action variables than metal-rich stars. Additionally, the metal-poor stars display decreasing trends in $\sigma_{J_r}$ and $\sigma_{J_z}$, while the metal-rich stars do not show any clear trends in $\sigma_{J_r}$ and $\sigma_{J_z}$ with increasing metallicity. In the $\sigma_{J_{\phi}}$-[Fe/H] plane, metal-poor stars remain relatively constant at $\sim 250 \, \text{kpc}\ \text{km/s}$, while a downward trend is observed in metal-rich stars as metallicity increases. The varying large action dispersions imply that the metal-poor stars in the Galactic Bulge have different dynamical origins (the first row of Figure~\ref{sigma}). We refer to the metal-poor stars as ``kinematically incoherent'', while the metal-rich stars exhibit similar kinematic patterns, possibly sharing a common origin that is distinct from that of the metal-poor stars \citep{Matteucci2019}, making them ``kinematically coherent''.

We use GMM to identify different ``kinematic populations" among metal-poor stars. $J_r$, $J_{\phi}$, and [Fe/H] are chosen as input parameters because of their relatively broad ranges and the clear functional relationship between $\sigma_r$ and [Fe/H]. The vertical action $J_z$ is omitted due to its small dispersions for both metal-rich and metal-poor stars, which may not provide sufficient distinction when using GMM. However, the BIC values begin to fluctuate as the number of Gaussian components increases. Consequently, minimizing the BIC value to determine the optimal number of components may not be appropriate for this clustering process. The Silhouette Coefficient (SC) is used to determine the optimal number of components, which is defined as follows:
\begin{equation}
    S(i) = \frac{b(i)-a(i)}{\text{max}\{a(i),b(i)\}}
\end{equation}
\begin{equation}
    \text{SC} = \frac{1}{n}\sum_{i=1}^{n}S(i)
\end{equation}

where $a(i)$ is the mean distance between a data point $i$ and other points in the same cluster, and $b(i)$ is the mean distance between the data point $i$ and points in other clusters. As a result, $S(i)$ stands for the silhouette coefficient for a single data point $i$ and SC is the silhouette coefficient for a whole cluster. The value of SC ranges from $-1$ to $1$, with higher values indicating better performance of GMM for a given number of components. By adopting SC criterion, the optimal number of Gaussian components is 3 for metal-poor stars with an SC value of 0.412, and 2 for metal-rich stars with a surprising SC value of 0.999. However, in the two clusters of metal-rich stars, one cluster contains only a single star, likely an outlier. This suggests that the entire metal-rich sample can be regarded as one coherent cluster, confirming that metal-rich stars are kinematically coherent and show no metallicity bias. For the clustering results of metal-poor stars, we label them as Cluster 0, Cluster 1 and Cluster 2.

The Wasserstein distance was employed to assess the similarity among three clusters based on three input parameters (i.e., $J_r$,$J_z$ and [Fe/H]). The distance is defined as $W_p(P, Q) = \left( \inf_{\gamma \in \Gamma(P, Q)} \int_{X \times X} d(x, y)^p \, \mathrm{d}\gamma(x, y) \right)^{1/p}$, which measures the minimum average cost required to transport mass from distribution \textit{P} to distribution \textit{Q}. Therefore, a smaller distance indicates greater similarity between the two distributions. After normalization, one minus the normalized values was used to determine the distribution similarity of the three clusters under three output parameters, producing a heatmap. As shown in Figure~\ref{heatmap}, Clusters 0 and 2 exhibit similar distributions in terms of $J_r$ and [Fe/H], with similarities of 0.80 and 0.68, respectively. However, the two clusters differ significantly under the $J_{\phi}$ distribution, suggesting that the metal-poor stars are well-clustered according to their features by GMM. The rightmost panel in Figure~\ref{heatmap} illustrates the distributions of the results in $J_r / J_{\text{total}}$-$|J_{\phi}| / J_{\text{total}}$ plane. Cluster 2 is the least metal-poor and is primarily supported by rotational motion, which aligns with previous findings indicating that the rotational signal increases with higher [Fe/H] \citep{Ness2016,Arentsen2020}. By contrast, Cluster 1 is the most metal-poor and has more stars with radial motions. The more metal-poor stars tend to be older and are perturbed by complex gravitational interaction at a longer timescale or perhaps experience more stellar encounters \citep{McTier2020}, thus it is possible that the stars in cluster 1 have more eccentric orbits than stars in Cluster 2. Cluster 0 may be composed of smaller groups due to its wide distribution of stars. Notably, the mean values of [Fe/H] for Cluster 0, Cluster 1 and Cluster 2 are $-0.40$, $-0.52$ and $-0.35$, aligning with the peaks found in the MDF (see Table~\ref{mdf}). However, the multiple peaks in the metal-rich range do not exhibit any kinematic differences as suggested by GMM. The possible origins of metal-rich stars will be analyzed with precise elemental abundances in subsequent analyses.

\subsection{The Retrograde Stars in the Galactic Bulge}
\begin{figure*}
    \begin{center}
        \includegraphics[width=\textwidth]{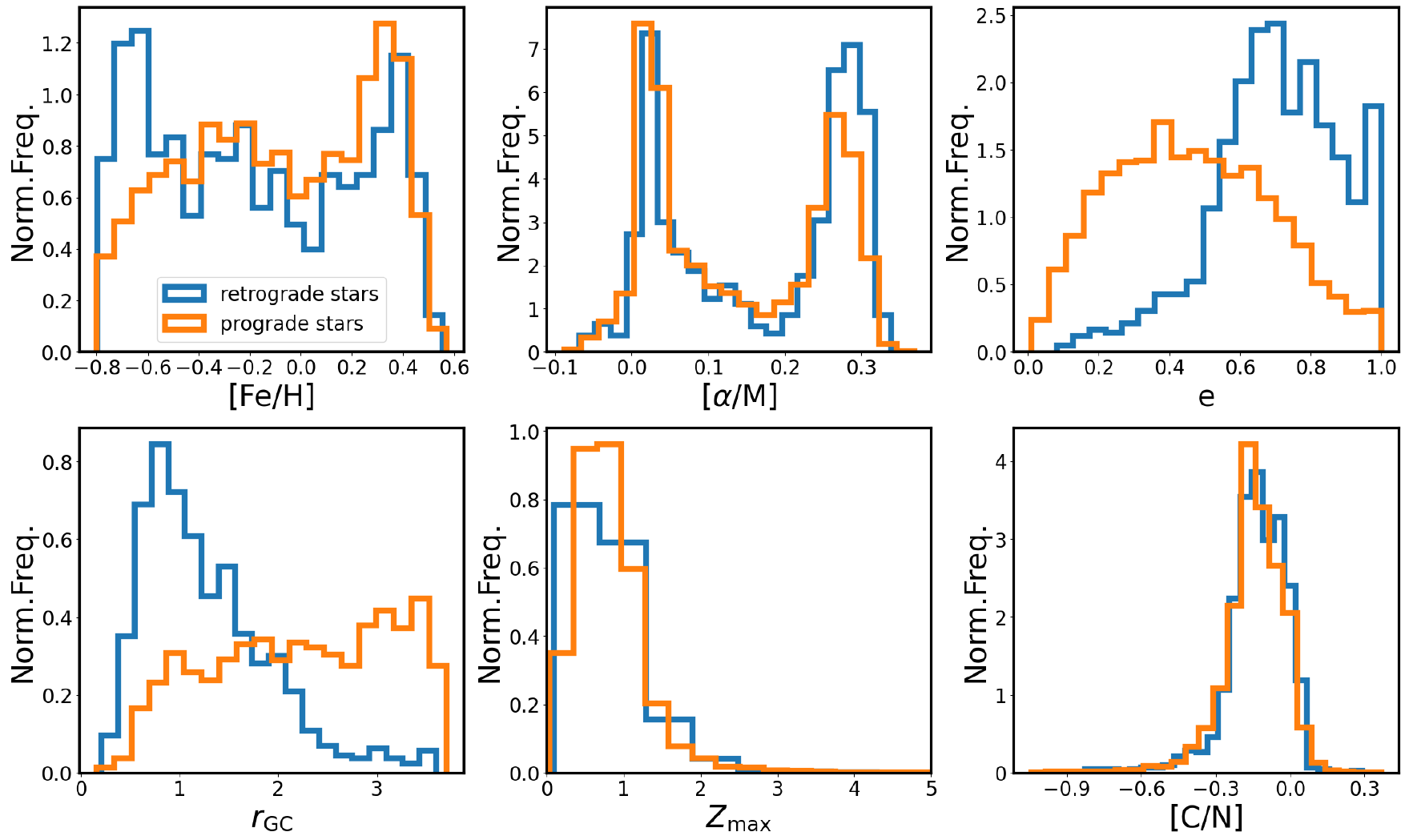}  
        \caption{The histograms of retrograde and prograde stars are presented in terms of elemental abundances and orbital parameters. The retrograde stars show a metal-poor peak at $\sim-0.6$ and are more metal-poor than prograde stars. The retrograde stars are also slightly [C/N] and $\alpha$ enhanced than prograde stars, indicating that they are relatively older. The retrograde stars are mainly confined within $r_{\text{GC}}\sim2$ kpc while the prograde stars have a large $r_{\text{GC}}$ more than 3 kpc. Both retrograde and prograde stars have $Z_{\text{max}}<3$ kpc.}
        \label{ret}
    \end{center}
\end{figure*}

\begin{figure}
    \begin{center}
        \includegraphics[width=0.5\textwidth]{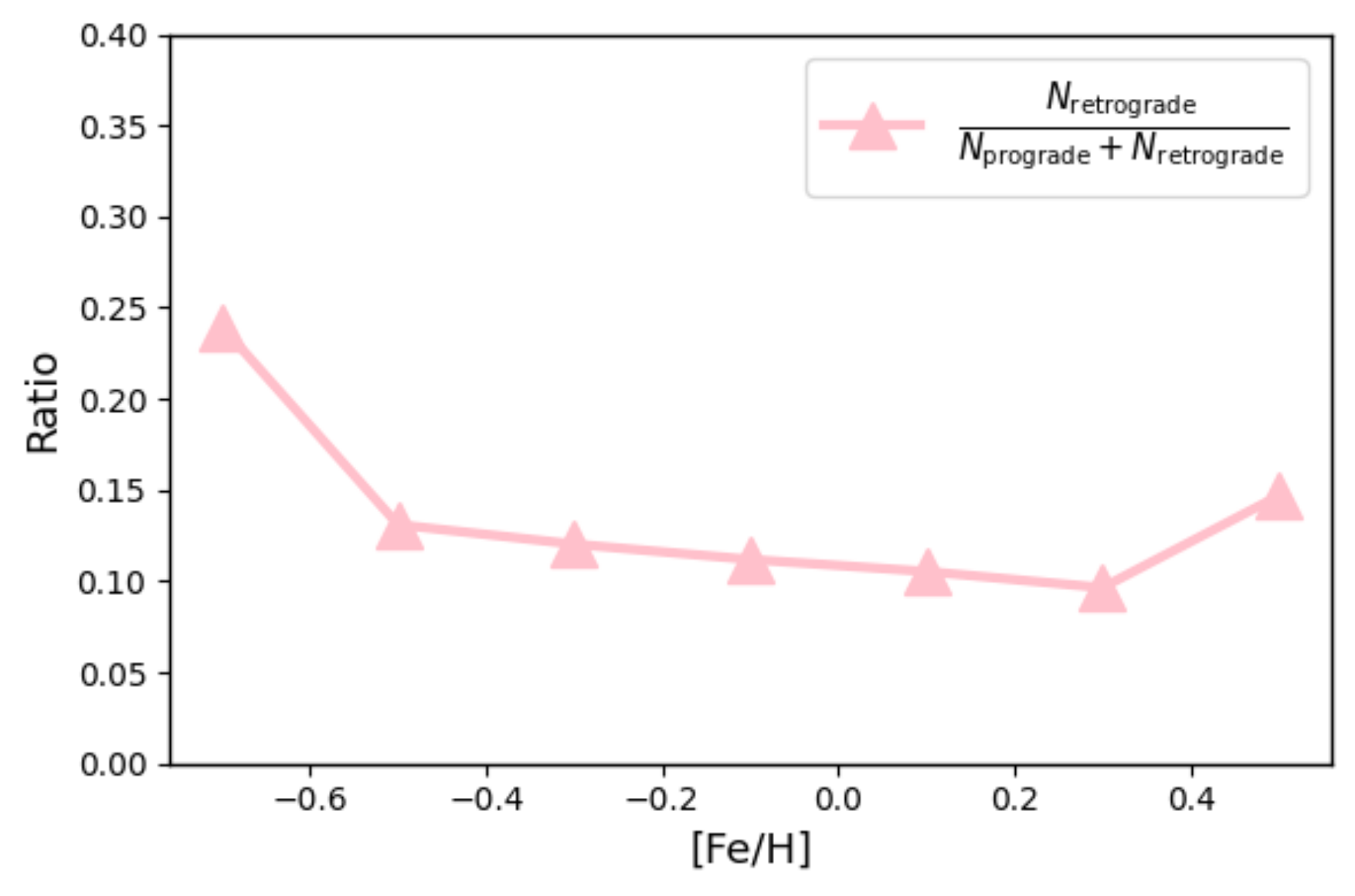}  
        \caption{The ratio of the number of retrograde stars to the total number of stars as a function of metallicity.}
        \label{ratio}
    \end{center}
\end{figure}

The retrograde stars ($V_{\phi}<0$) are confined within \( r_{\text{GC}} \sim 1.11_{-0.45}^{+0.80} \) kpc of the Bulge across all metallicities (see bottom left panel of Figure~\ref{ret}), compared to prograde stars, which are found at \( \sim 2.28_{-1.14}^{+0.98} \) kpc. This observation is noteworthy, and we briefly discuss the properties of these stars here. In Figure~\ref{ratio}, the fraction of retrograde stars in each [Fe/H] bin decreases with increasing metallicity (except for the very metal-rich end). \citet{Berokurov2022} found that the Galactic disk experienced a process called ``spin-up'' due to the rapid formation of the MW disk over $\sim 1-2$ Gyr, during which the median $V_{\phi}$ of the in-situ stars increases sharply with increasing metallicity between $-1.3$ and $-0.9$. Although our Bulge samples are limited to [Fe/H] $>-0.8$, there is still an evolution in the $V_{\phi}$-[Fe/H] plane for [Fe/H] $>-0.8$ \citep[see Figure 4 in ][]{Berokurov2022}, which possibly explains the decreasing ratio trend if the Bulge formed from the disk, such as a clumpy disk origin \citep{Inoue2012}. The high ratio at the very metal-rich end could be attributed to either the fewer data points causing larger scatter, or these very metal-rich stars have a much more complex dynamical origin.


\citet{Queiroz2021} also identified those retrograde stars (which they defined as $V_{\phi}<-$50 km/s) using APOGEE DR16 and found that the retrograde stars were more metal-poor and more eccentric than prograde stars (see Figure~\ref{ret}). They speculated that these retrograde stars could be the counterparts of the Splash debris \citep{Belokurov2020} but contaminated by metal-rich stars, which consists of proto-disk stars heated by the last major merger, featuring [Fe/H]$>-0.7$ and highly eccentric characteristics. Additionally, the retrograde stars are slightly more [C/N] and $\alpha$ enhanced than the prograde stars, suggesting that they are relatively older 
\citep{Salaris2015,Paula2021}. However, the Splash stars have mean cylindrical radii of $6 \sim 9$ kpc in the solar vicinity, making it difficult to explain why Splash counterparts are observed right in the Bulge. Both retrograde metal-poor and metal-rich stars have $V_R$ dispersions of $\sim116$ km/s, which is only about half that of the Splash stars, according to the simulation \citep{Liao2024}. Consequently, we exclude the ``Splash counterpart'' hypothesis and instead propose that these stars are very likely born in situ in the Bulge, given their small galactocentric distances, or that they are from the migration of disk stars with a clumpy origin \citep{Amarante2020}.

\begin{figure*}
\sidecaption
  \includegraphics[width=12cm]{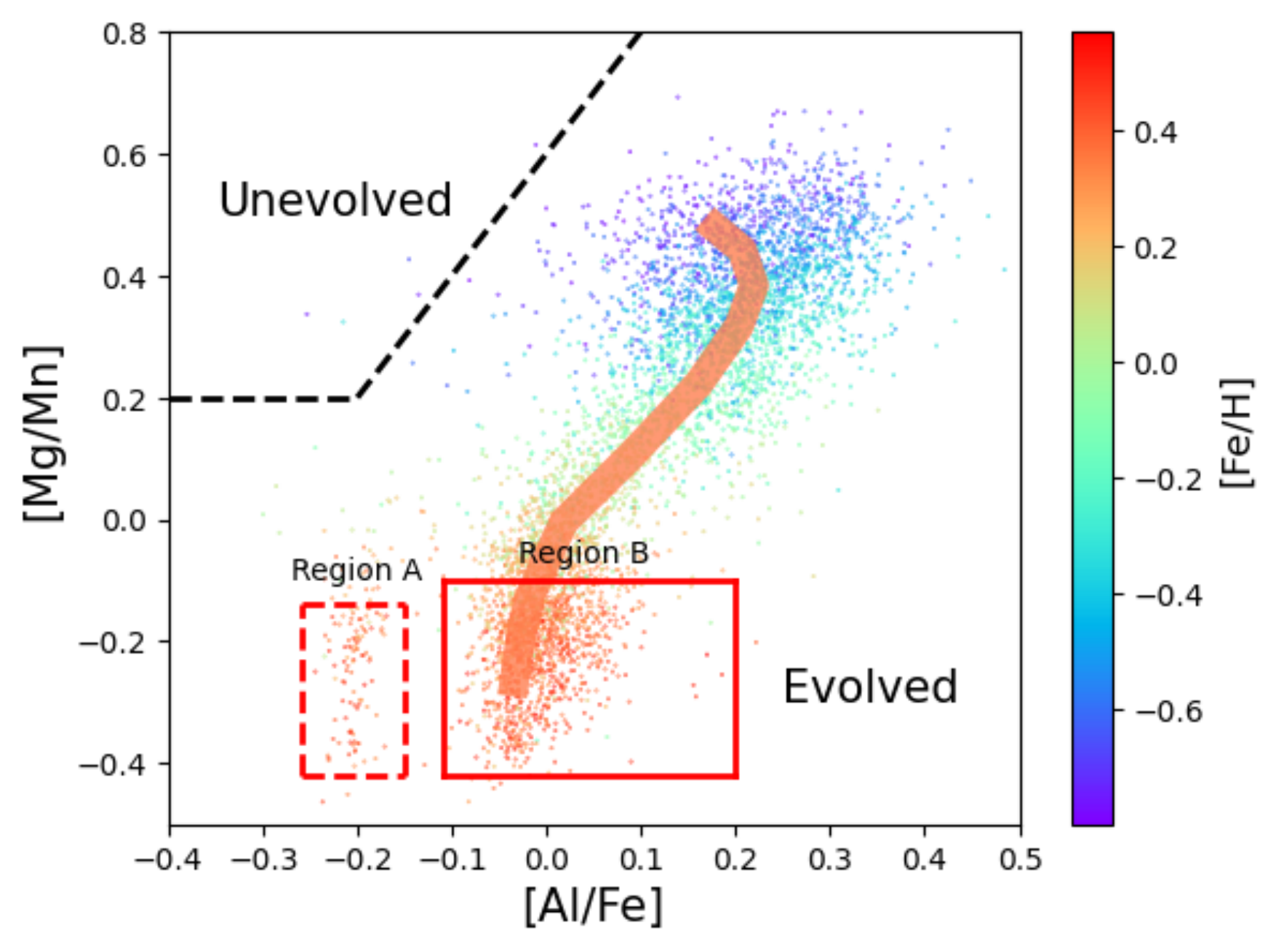}
     \caption{The distributions of bulge stars in the [Mg/Mn] - [Al/Fe] plane show that there are almost no accreted stars in the unevolved region. The coral-colored wide line represents the evolutionary path that fits the observations. The metal-rich stars in region A and region B show a clear gap in [Al/Fe] values, suggesting the low [Al/Fe] metal-rich stars in region A could have an accretion origin.}
     \label{mg}
\end{figure*}

\subsubsection{The [Mg/Mn]-[Al/Fe] Plane}
The [Mg/Mn]-[Al/Fe] plane is used to chemically select the accreted populations \citep{HawKins2015,Das2020}. The accreted and in-situ regions are also referred to as ``unevolved'' and ``evolved'' regions, respectively (see \citet{Horta2021} for more details). The MW and its satellites follow different evolutionary paths in this plane, influenced by their initial mass functions and masses. Typically, the evolutionary paths of satellites end at lower [Al/Fe] values, whereas the MW reaches higher [Al/Fe] values \citep{Fernandes2023}. In the [Mg/Mn] - [Al/Fe] plane (see Figure~\ref{mg}), we notice almost no accreted populations in the unevolved region. This is not surprising, as the inner Galactic substructure, such as Heracles, is very metal-poor, with a mean [Fe/H] of $-1.26$. In the vicinity of regions A and B, there is a distinct gap of [Al/Fe] values between the metal-rich stars found in these two regions. The metal-rich stars in region A have [Al/Fe]$<-0.15$, while the metal-rich stars in region B have [Al/Fe]$>-0.10$. As can be seen in Figure~\ref{mg}, the low-Al stars are well-separated from the rest distributions with a clear gap.
According to \citet{Fernandes2023}, dwarf galaxies typically end their evolutionary paths in region A, while larger galaxies like the MW, conclude in region B. Therefore, it is possible that the stars in region A are accreted, indicating a recent merger event (i.e., Virgo Radial Merger; Donlon et al.\citeyear{Donlon2019}), as these stars are at the end of their evolutionary path. \citet{Feuillet2022} also identified these low [Al/Fe] (defined as [Al/Fe]$<-0.20$) metal-rich stars in the Galactic disk using APOGEE DR17 and found they were extremely old, thus suggesting an accretion origin.

\begin{figure*}
    \begin{center}
        \includegraphics[width=\textwidth]{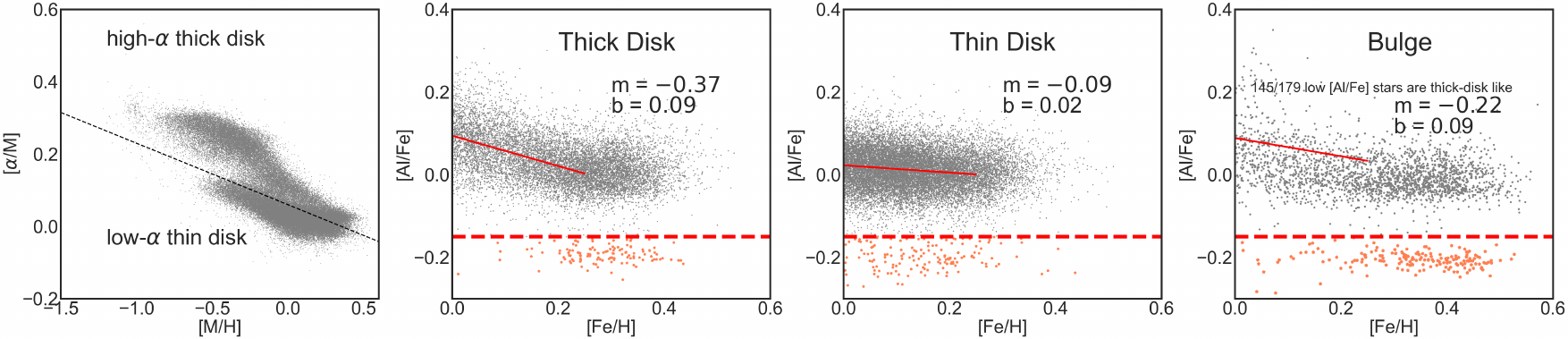}  
        \caption{Leftmost panel: the disk samples with $3.5<R<8$ kpc and $0<|z|<1$ kpc without metal-poor stellar halo contamination removed. The right three panels: the distributions in [Al/Fe]-[Fe/H] plane of thick disk, thin disk and the Bulge. The red dashed line indicates where [Al/Fe]=$-0.15$. The red solid line represents the linear fit for high Al (\([\text{Al}/\text{Fe}]\)) metal-rich stars when \(0 < \text{[Fe/H]} < 0.25\), where \(m\) is the slope and \(b\) is the intercept. 145 out of 179 low Al metal-rich Bulge stars exhibit thick disk-like chemical characteristics.}
        \label{disk}
    \end{center}
\end{figure*}

To study the correlation of these low [Al/Fe] metal-rich stars between in the bulge and the disk, we also select similar stars from the thick and thin disks for comparison, as shown in Figure~\ref{disk}. The disk sample stars are selected with the same quality cuts as the Bulge stars but with different spatial constraints: $3.5<R<8$ kpc and $0<|z|<1$ kpc. Here we also do not remove the contamination of the metal-poor halo stars, since we only consider the metal-rich stars. The high-$\alpha$ thick disk and low-$\alpha$ thin disk are separated according to the relation: [$\alpha$/M] = $-0.17*$[M/H] + 0.06. We define low-Al metal-rich stars with [Al/Fe]$<-0.15$ and [Fe/H]$>0$. 
From Figure~\ref{disk}, we observe that the thick disk still exhibits [Al/Fe] evolution with respect to [Fe/H] in the range of 0 to 0.25, showing a slope of $-0.36$ from a linear fit of the [Al/Fe]-[Fe/H] plane. By contrast, the thin disk shows a mild evolution with a slope of $-0.09$, while the Bulge exhibits a slope of $-0.22$, confirming the previous finding that there are chemical similarities between the thick disk and the Bulge \citep{Alves2010}. Moreover, \citet{Di2016} found that the chemistry of the Bulge and Galactic disks became evident when considering the radial and vertical structure and the thick disk was essential for a whole picture of the Bulge \citep[also see][]{Rojas2017}. Among the 179 low Al metal-rich stars in the Bulge, 145 ($\sim80$\%) exhibit a thick disk-like relation, having [$\alpha$/M] greater than $-0.17*$[M/H] + 0.06. As a result, it is likely that these low Al metal-rich stars are associated with those found by \citet{Feuillet2022}.

Combining the previous analysis on kinematics with this work, it is very possible that the metal-rich stars in the Bulge have a disk origin due to migration caused by the growing bar. However, the metal-rich stars are not all contributed by the thin disk as \citet{Boin2024} suggested. The exact origin of the low Al metal-rich stars is difficult to determine due to the mass-metallicity relation \citep{Brinchmann2004}.

\begin{figure*}
    \begin{center}
        \includegraphics[width=\textwidth]{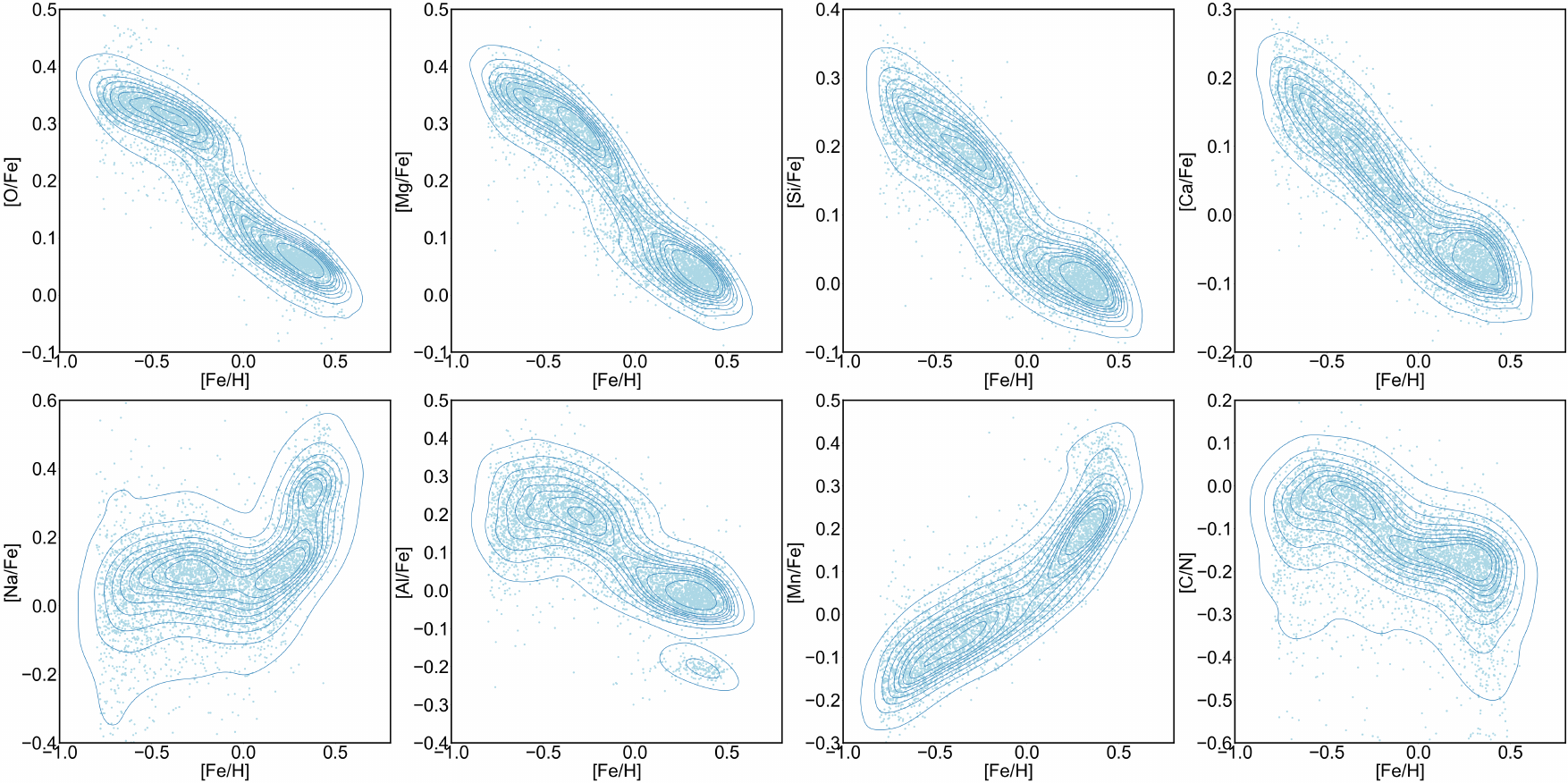}  
        \caption{The elemental distributions of the Bulge in terms of  O, Mg, Si, Ca, Na, Al, Mn elements and [C/N] ratio, with density contours plotted by Kernel Density Estimation (KDE). Notably, the low Al metal-rich stars are also identified.}
        \label{element}
    \end{center}
\end{figure*}

\begin{figure*}
\sidecaption
  \includegraphics[width=12cm]{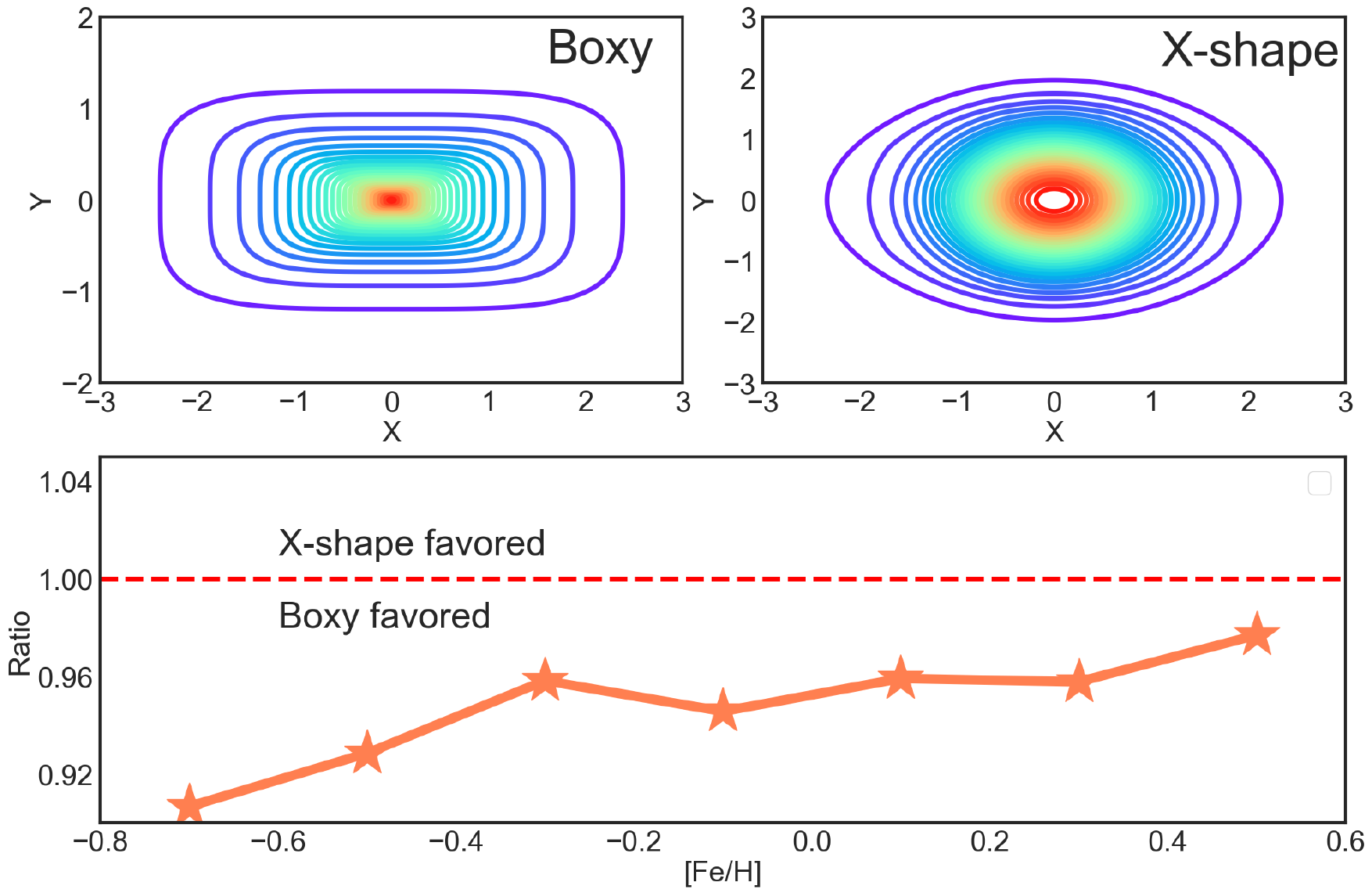}
     \caption{Top row: the density functions projected in X-Y plane with Z = 0, where $\rho_0$ is chosen to be 1. Bottom row: the relative possibility of supporting a certain profile as a function of metallicity. The red dashed line indicates the same possibility of supporting two profiles. For each [Fe/H] bin, there are 15 bins in the X direction, 15 bins in the Y direction and 3 bins in the Z direction.}
     \label{structure}
\end{figure*}

\subsubsection{The Elemental Abundance Distributions }
Figure~\ref{element} shows the chemical distributions of $\alpha$ elements (O, Mg, Si, Ca), odd-z elements (Na, Al), iron-peak element (Mn) and [C/N] ratios. The existing bimodality of [$\alpha$/Fe]-[Fe/H] and [C/N]-[Fe/H] distributions confirms the former observations \citep{Rojas2019,Queiroz2021}. The knees of the \([\alpha/\text{Fe}]\)–[\text{Fe/H}] planes (where \(\alpha\) denotes O, Mg, Si, and Ca) are all at \([\text{Fe/H}] \sim -0.2\), which is higher than previous studies \citep{Hill2011,Friaca2017} and also higher than that of thick disk stars \citep{Bensby2017,Rojas2017}. This suggests that the star formation rate in the Bulge could be higher than previously considered, as there are more Type Ia supernovae dominating over Type II supernovae at later times \citep{Mason2024} as well as a shorter formation timescale \citep{Lecureur2007}. The production of odd-z elements is metallicity-dependent and depends on the surplus of neutrons from Ne made in the CNO cycle \citep{Kobayashi2020}. As shown in Figure~\ref{element}, the [Al/Fe]-[Fe/H] distribution also exhibits a bimodality \citep{McWilliam2016},  with low Al metal-rich stars identified through Gaussian kernel estimation and a similar knee to those of the $\alpha$ elements. In contrast, the distribution of [Na/Fe]-[Fe/H] does not align with the Bulge evolution model proposed by \citet{Kobayashi2006} with a flat Initial Mass Function (IMF), while [Mn/Fe]-[Fe/H] distribution is in agreement with the model.

The observed [Na/Fe] shows a flat trend when [Fe/H]$<0$ and rises with increasing [Fe/H] in the super-solar range. The Na element is produced in stars of various masses. For massive stars, carbon burning is the main source \citep{Woosley1995}, while neon-sodium (NeNa) cycle is the main source for stars with lower masses. Na could also be produced through Type Ia supernovae. One possible explanation for the observed increase in [Na/Fe] at higher metallicities is that more massive stars are producing Na more efficiently, leading to fewer white dwarfs and thus allowing Na to be enriched more rapidly than Fe over time. The steep rise of [Na/Fe] exhibits a well-behaved, metallicity-dependent sodium yield, although the Na abundance estimates for metal-rich stars may be overestimated due to saturation effects.
We also notice that there are stars with lower [C/N] abundance across all metallicities, which suggests that both metal-poor and metal-rich stars contain an adequate number of young stars, implying a complexity of Bulge stellar populations.

\subsection{The Structural Dependence on Metallicity in the Galactic Bulge}
 The B/P shape of the Bulge has been studied through RCG star counts method \citep{Stanek1994,Ba2005,NessL2016}. The X-shape of the bulge has also been identified through the ``split in the red clump'' observed in star counts along the line of sight toward the bulge \citep{NessL2016}. However, there are different viewpoints suggesting that this X-shape may be an artifact resulting from the split in RCG stars due to the presence of different stellar populations \citep{Lopez2016,Lopez2019}. Some simulations \citep{A2017, semczuk2022} have revealed that the X-shape is most prominent in more metal-rich or younger stars, while a boxy or spheroidal component is clearly observed among more metal-poor or older stars. To avoid confusion, it should be clarified that both the boxy/peanut structure and the X-shape are manifestations of the same structural feature in the Bulge—the X-shape is essentially a spatial extension of the boxy/peanut structure, or more precisely, the X-shape is ``embedded'' within the boxy/peanut configuration \citep[see Figure 3 and Figure 4 of][]{A2017}. Moreover, the boxy and spheroid distributions can be related, representing hotter populations according to \cite{Debattista2017}.
 
 To quantitatively study the structural dependence on metallicity in the Galactic Bulge, we adopt the density functions for both the boxy and X-shaped profiles from \citet{Lopez2005} and \citet{wegg2013} respectively: 

 \begin{equation}
     \rho_{\text{boxy}} = \rho_{0}\exp{\left(-\frac{(x^4 + (\frac{y}{0.25})^4 + (\frac{z}{0.4})^4)^\frac{1}{4}}{740 \text{pc}}\right)}
 \end{equation}

\begin{equation}
\begin{split}
    \rho_{\text{X-shape}} = \rho_{0} & \exp{\left(-\frac{s_1}{700\text{pc}}\right)}\exp{\left(-\frac{|z|}{322\text{pc}}\right)} \\
    & \times \left(1 + 3\exp{\left(-\left(\frac{s_2}{1000\text{pc}}\right)^2\right)} \right. \\
    & \left. + \exp{\left(-\left(\frac{s_3}{1000\text{pc}}\right)^2\right)}\right)
\end{split}
\end{equation}
where $s_1 = \text{Max}\left[2100\text{pc}, \sqrt{x^2 + \frac{y}{0.7}^2}\right]$, $s_2 = \sqrt{(x - 1.5z)^2 + y^2}$, $s_3 = \sqrt{(x + 1.5z)^2 + y^2}$. The parametrization is provided by \citet{Lopez2017}. Following \citet{c2022}, we minimize the $\chi^2$ to fit the two models, which is defined as follows:
\begin{equation}
    \chi^2 = \sum^N_{i=1}\frac{|\rho_{i,\text{model}}-\rho_{i,\text{data}}|}{\sigma_i^2}
\end{equation}
$\sigma_i = \sqrt{\rho_{i,\text{data}}\rho_1}$, where $\rho_1$ corresponds to the Poisson error of each spatial bin and $\rho_1$ is inversely proportional to the bin size. When $\rho_{i,\text{data}} = 0$, $\sigma_i = \rho_1$. We divide the data into 26 bins in the X direction, 26 bins in the Y direction and 10 bins in the Z direction. We find that the spatial distribution of the data supports a boxy profile more than an X-shape profile with a smaller $\chi^2$ value. Here, we do not calculate the corresponding probabilities from the cumulative distribution function, as the probabilities vary significantly with different bin sizes. After testing with different bin sizes, we find that the Bulge data generally more support the boxy profile with smaller $\chi^2$ values.

We use the ratio $\chi^2_{\text{X-shape}}/\chi^2_{\text{boxy}}$ as the relative probability of supporting a certain profile, as a function of [Fe/H] as well as stellar age. As shown in Figure~\ref{structure}, the observed data across all metallicities support a boxy profile. However, the ratio exhibits an upward trend with increasing metallicity, indicating that metal-rich stars are more likely to support an X-shaped profile than metal-poor stars, which is in agreement with the simulation from \citet{A2017}. However, in the study by \citet{Zoccali2018}, metal-rich stars exhibit a distinct boxy profile when observed from the Sun's perspective, which contradicts the simulation from \citet{A2017}. The discrepancy may arise because the simulation introduced a major merger event $8-10$ Gyr ago, rather than allowing the Bulge to evolve secularly driven by bar instability. From Figure~\ref{structure}, although metal-rich stars show a higher tendency toward X-shape structures, they still predominantly exhibit a boxy profile. This suggests that while the major merger may influence the spatial distribution, it is likely not the dominant factor.

It needs to be noted that we do not include comparisons with other survey data or the sample selection effects, as these are beyond the scope of this work. However, we can still clearly observe how the structure of the Bulge depends on metallicity and stellar age, which has not been extensively discussed using observational data.

\section{Discussion and Summary}\label{summary}
Using elemental abundances from APOGEE DR17 and astrometric data from \textit{Gaia} DR3, we examined how kinematics, elemental abundances and spatial distribution of  bulge stars correlate with metallicity. We confirmed the trend of line-of-sight velocity dispersions along the minor axis of the Bulge, noting that metal-rich stars exhibit this trend with a turning point at \( |b| \sim 6.5^\circ \), while metal-poor stars do not. This indicates that metal-rich stars are trapped by bar instability, whereas metal-poor stars have a more complex dynamical origin. In the MDF of Bulge stars, we identified six peaks, with the most metal-rich peak showing the highest prominence. Metal-poor stars exhibit larger dispersions in actions compared to metal-rich stars, indicating that metal-poor stars are kinematically incoherent, while metal-rich stars are relatively coherent. This suggests that metal-poor stars may have multiple dynamical origins. 

In addition, we applied GMM in ($J_r$,$J_{\phi}$ and [Fe/H]) space for both metal-poor and metal-rich stars. According to the SC estimation criterion, metal-rich stars are better represented by a single cluster, suggesting a shared dynamical origin. In contrast, at least three distinct clusters are identified among the metal-poor stars. We also noticed that the retrograde stars are centered within $\sim 1.11_{-0.45}^{+0.80}$ kpc of the Bulge. Generally, more metal-rich stars exhibit a lower retrograde ratio, which can be interpreted as evidence that the Bulge has a disk origin and that these stars experienced a ``spin-up'' process within the disk. These retrograde stars could also be in-situ stars without ``Splash'' origin if the Bulge has a clumpy disk origin \citep{Amarante2020}.

In the [Mg/Mn]-[Al/Fe] plane of Bulge sample stars, we observe that a few metal-rich stars with low Al abundances may be associated with accreted substructures in the disk \citep{Feuillet2022}. If this is the case, $\sim$80\% of these stars exhibit a thick disk-like $\alpha$-abundance pattern, and they may have migrated from the thick disk due to the growing bar. However, it is difficult to determine whether these stars are indeed accreted, as they are too metal-rich to originate from typical dwarf galaxies. The elemental distributions of Bulge stars are generally aligned with the theoretical model proposed by \citet{Kobayashi2006}. The sodium-enhanced and metal-rich stars are well represented in our samples, indicating that there could be more massive stars in the Bulge than we previously considered, requiring a much flatter IMF or probably an IMF in logarithmic normal form \citep[e.g., see the comparison of Figure 21 from][]{Johnson2014}. In the previous study, \citet{Lecureur2007} has found that Bulge stars have higher O, Mg and Al abundances than disk stars when [Fe/H]$>-0.5$, which also suggests a possible top-heavy IMF for the Galactic Bulge \citep[also see][]{C2011}. We also briefly studied the spatial distribution dependence on metallicity, and found that across all metallicities, the stars have a higher probability of supporting a boxy profile of the Bulge. Metal-rich stars show higher probabilities of exhibiting an X-shaped profile compared to metal-poor stars; however, they still generally support a boxy profile.

In summary, the origins of metal-poor stars still need further exploration through future surveys. Conversely, the metal-rich stars are similar in kinematics, and there is clear evidence that they have a disk origin. However, the metal-rich stars could not only originate from the thin disk, as simulations suggest, but also from the thick disk.

\begin{acknowledgements}
We thank the anonymous referee for insightful suggestions that have greatly improved this manuscript. This work was supported by National Key R\&D Program of China No. 2024YFA1611900, and the National Natural Science Foundation of China (NSFC Nos. 11973042, 11973052). Funding for the Sloan Digital Sky Survey IV has been provided by the Alfred P. Sloan Foundation, the U.S. Department of Energy Office of Science, and the Participating Institutions.

SDSS-IV acknowledges support and resources from the Center for High Performance Computing  at the University of Utah. The SDSS website is \url{www.sdss4.org}.

SDSS-IV is managed by the Astrophysical Research Consortium for the Participating Institutions of the SDSS Collaboration including the Brazilian Participation Group, the Carnegie Institution for Science, Carnegie Mellon University, Center for Astrophysics | Harvard \& Smithsonian, the Chilean Participation Group, the French Participation Group, Instituto de Astrof\'isica de Canarias, The Johns Hopkins University, Kavli Institute for the Physics and Mathematics of the Universe (IPMU) / University of Tokyo, the Korean Participation Group, Lawrence Berkeley National Laboratory, Leibniz Institut f\"ur Astrophysik Potsdam (AIP),  Max-Planck-Institut 
f\"ur Astronomie (MPIA Heidelberg), Max-Planck-Institut f\"ur Astrophysik (MPA Garching), Max-Planck-Institut f\"ur Extraterrestrische Physik (MPE), National Astronomical Observatories of China, New Mexico State University, New York University, University of Notre Dame, Observat\'ario Nacional / MCTI, The Ohio State University,Pennsylvania State University, Shanghai Astronomical Observatory, United 
Kingdom Participation Group, Universidad Nacional Aut\'onoma de M\'exico, University of Arizona, University of Colorado Boulder, University of Oxford, University of Portsmouth, University of Utah, University of Virginia, University of Washington, University of Wisconsin, Vanderbilt University, and Yale University.

This work has made use of data from the European Space Agency (ESA) mission
{\it Gaia} (\url{https://www.cosmos.esa.int/gaia}), processed by the {\it Gaia}
Data Processing and Analysis Consortium (DPAC,
\url{https://www.cosmos.esa.int/web/gaia/dpac/consortium}). Funding for the DPAC
has been provided by national institutions, in particular the institutions
participating in the {\it Gaia} Multilateral Agreement  
\end{acknowledgements}

%
%

\bibliographystyle{aa}
\bibliography{liu}{}
\end{document}